\documentclass[aip,amsmath,amssymb,%
 reprint%
]{revtex4-1}

\usepackage[utf8]{inputenc}

\usepackage{textgreek} %
\usepackage{verbatim}

\usepackage{multirow}

\usepackage{color}

\usepackage{graphicx}%
\usepackage{dcolumn}%
\usepackage{bm}%
\usepackage[utf8]{inputenc}
\usepackage[T1]{fontenc}
\usepackage{mathptmx}

\definecolor{napiergreen}{rgb}{0.16, 0.5, 0.0}
\begin{document}

    \title[Markov state models of RNA fraying]{
The mechanism of RNA base fraying: 
 molecular dynamics simulations analyzed with core-set Markov state models
}
\author{Giovanni Pinamonti}
    \email{giovanni.pinamonti@fu-berlin.de}
    \affiliation{Department for Mathematics and Computer Science, Freie Universit{\"a}t, Berlin, Germany}
    \author{Fabian Paul}
    \affiliation{Department of Biochemistry and Molecular Biology, Gordon Center for Integrative Science,
      The University of Chicago, Chicago, IL 60637}
    \author{Frank No\'e}
    \affiliation{Department for Mathematics and Computer Science, Freie Universit{\"a}t, Berlin, Germany}
    \author{Alex Rodriguez}
    \affiliation{ICTP, International Centre for Theoretical Physics, Trieste, Italy}
    \author{Giovanni Bussi}
    \email{bussi@sissa.it}
    \affiliation{Scuola Internazionale Superiore di Studi Avanzati, via Bonomea 265, Trieste, Italy}

\begin{abstract}%
The process of RNA base fraying (i.e. the transient opening of the termini of a helix) is involved in many aspects of RNA dynamics.
We here use molecular dynamics simulations and Markov state models to characterize
the kinetics of RNA fraying and its sequence and direction dependence.
In particular, we first introduce a method for determining biomolecular dynamics employing core-set Markov state models
constructed using an advanced clustering technique.
The method is validated on previously reported simulations. We then use the method to analyze extensive trajectories
for four different RNA model duplexes.
Results obtained using D. E. Shaw research and AMBER force fields are compared and discussed in detail, and show a non-trivial interplay between
the stability of intermediate states and the overall fraying kinetics.
\end{abstract}

    \maketitle

\section{Introduction}

Ribonucleic acid (RNA)
plays a fundamental role in the biology of the cell.\cite{morris2014riseregRNA}
 RNA molecules fold in intricate structures, that undergo complex rearrangements,\cite{al2008rna}
 to fulfill a number of
  biological functions, such as gene regulation, splicing, catalysis, and protein synthesis.
  It is thus key to get a more precise understanding of the mechanisms involved in RNA folding and conformational transitions.
  Current experimental techniques are limited to ensemble measurements or to low spatiotemporal resolution.
 For this reason, computational tools are fundamental for the study of biomolecular
 systems, including ribonucleic acids.
 Molecular dynamics (MD) simulations using empirical force fields, propelled by numerous
 theoretical and technical improvements,\cite{bowman2013introduction,shaw2014anton,valsson2016enhancing,vojtvech2018exploring,camilloni2018advanced} have enabled scientists
 to accurately study the thermodynamics and kinetics of proteins\cite{klepeis2009long,lane2013milliseconds,preto2014fast,plattner2017complete} and nucleic acids.\cite{vangaveti2017advances,smith2017physics,sponer2018rna,dans2018modeling}
 In particular, the framework of Markov state models (MSMs)\cite{schutte1999direct,swope2004describing,noe2007hierarchical,chodera2007automatic,prinz2011markov} makes it possible
 to perform a systematic analyses of the metastable states and kinetics of biomolecular systems.
 In principle, these computational tools could be used as a highly accurate ``computational microscope'',\cite{lee2009discovery} allowing the quantitative description
 of the individual steps leading to, \emph{e.g.}, the rupture and formation of a double helix.
 In practice, the results that can be obtained for RNA molecules are still limited
 by several factors, the most important of which being the accuracy of the force fields employed.
\cite{bergonzo2015highly,kuhrova2016computer,bottaro2016free,sponer2018rna}

  The process of ``base fraying'',
 that is the breaking of base pairing and stacking interactions at the termini of a RNA (or DNA) double helix,
is an apparently simple yet far from trivial process.
Frayed states are intermediate in the RNA zipping and unzipping processes,
have been proposed to be important in the interaction of RNA with proteins (see, \emph{e.g.},
Refs.~\onlinecite{betterton2005opening,sydow2009structural,colizzi2012rna,da2016bridge}),
and might be relevant in strand invasion~\cite{huang2013impact} and, in general,
in secondary structure rearrangements required for riboswitch function.\cite{serganov2013decade}
The characterization of fraying kinetics by experimental techniques is however difficult
due to their short lifetimes.\cite{snoussi2001imino,liu2008dynamic}
Base fraying in RNA  has been characterized by means of computer simulations in
several works.\cite{colizzi2012rna,zgarbova2014base,xu2016understanding}
Colizzi and Bussi \cite{colizzi2012rna} characterized the thermodynamics of the process,
suggesting that dangling bases at the 3'-end are more stable than those at the 5'-end and thus might be
important intermediates in duplex unzipping.
This finding is in agreement with the higher stabilization provided to duplexes by
3'-end dangling bases when compared to 5'-end dangling bases \cite{xia1998thermodynamic}, although this latter quantity also depends
on the energy of stacking in single stranded RNAs.
The stabilities computed by Colizzi and Bussi were however largely overestimated by the adopted unidirectional pulling,
and were thus only usable to rank the fraying propensity of different sequences.
Zgarbova et al \cite{zgarbova2014base} performed a detailed characterization of the non-canonical
structures observed with current force fields at the termini of DNA and RNA duplexes, without however aiming
at obtaining quantitative populations.
Both these works did not explicitly analyze the kinetics of the process.
Finally, Xu et al~\cite{xu2016understanding} %
presented a partial kinetic model reproducing the opening of the base on the 5' terminus of a RNA duplex.
However, a comparison of the kinetics of the two ends and a quantitative analysis of its sequence dependence are still missing.
A number of papers addressed fraying in DNA
(see, \emph{e.g.}, Refs.~\onlinecite{wong2008pathway,perez2010real,hagan2003atomistic}
and the already mentioned Ref.~\onlinecite{zgarbova2014base}) or in an RNA:DNA hybrid in complex with a protein.\cite{da2016bridge}

We here employ extensive MD simulations using 4 different sequences
with the goal of quantitatively characterizing the fraying kinetics, through the use of MSMs.
Specifically, we extend the core-based MSM framework~\cite{buchete2008coarse,schutte2011markov}
with density-based clustering.\cite{rodriguez2014clustering,d2018automatic}
The procedure is validated on the kinetics of short oligonucleotides first and then applied to base fraying in RNA duplexes.
We chose different sequences in order to assess the effect of the position
of purines/pyrimidines and the influence of the neighboring base pair.
Two state-of-the-art force fields are compared, specifically
i) the one recently published by the D. E. Shaw research laboratory~\cite{tan2018rna}
(in the following, DESRES)
and
ii) the latest refinement of the AMBER force field,\cite{banas2010performance}
which is the default AMBER force field for RNA systems.
Both force fields are based on previous versions of the AMBER force field.\cite{cornell1995second,perez2007refinement}
We studied the sequence dependence of stability, fraying rate, and the different pathways
which the process can follow.
Interestingly, the results display a non-trivial interplay between the stability of intermediate
structures and the kinetics of the process.

\section{Methods}

\subsection{Molecular dynamics simulations}

We simulated the dynamics of short helices composed of a 3 base-pair GC stem plus an AU terminal pair.
In particular, the following four permutations were used as model constructs:
$^{\texttt{5'-ACGC}}_{\texttt{3'-UGCG}}$,
$^{\texttt{5'-AGCG}}_{\texttt{3'-UCGC}}$,
$^{\texttt{5'-UCGC}}_{\texttt{3'-AGCG}}$,
$^{\texttt{5'-UGCG}}_{\texttt{3'-ACGC}}$.
In the rest of the paper we will refer to these constructs using only the sequence of the strand fraying at its 5'-end, respectively ACGC, AGCG, UCGC, UGCG.
Initial structures were obtained using the Make-NA server (\url{http://structure.usc.edu/make-na/}).
RNA duplexes were solvated in explicit water,
adding Na$^+$ counterions to neutralize the RNA
charge, plus additional NaCl to reach the nominal concentration of $0.1$~M.
The system was inserted in a truncated dodecahedral box with periodic boundary conditions and box size $5.17$ nm.
RNA was described using either the DESRES force field \cite{cornell1995second,perez2007refinement,tan2018rna} with TIP4P-D water model \cite{piana2015water}
or the AMBER force field \cite{cornell1995second,perez2007refinement,banas2010performance} with the TIP3P water model.\cite{jorgensen1983comparison}
Ions in DESRES simulations were described using the CHARMM parameters \cite{mackerell1998all} as recommended in Ref.~\onlinecite{tan2018rna},
whereas in AMBER simulations they were described using AMBER-adapted parameters for Na$^+$ \cite{aaqvist1990ion} and Cl$^-$.\cite{dang1995mechanism}
Based on previous results, we do not expect RNA dynamics to be highly affected by the ion parameters at this concentration.\cite{besseova2012simulations,pinamonti2015elastic}
The equations of motion were integrated with a $2$ fs time step.
All bond lengths were constrained using the LINCS algorithm.\cite{hess1997lincs}
Long-range electrostatics was treated using particle-mesh-Ewald summations.\cite{darden1993particle}
Trajectories were generated in the isothermal-isobaric
ensemble using stochastic velocity rescaling \cite{bussi2007canonical}
and the Parrinello-Rahman barostat.\cite{parrinello1981polymorphic}
All simulations were performed using GROMACS (version 4.6.7, calculations using AMBER force field, and 5.1.2, calculations using DESRES force field).
Force field parameters can be found at \url{https://github.com/srnas/ff}.
Since we decided to focus the study on the fraying of the A-U
terminal pair, we restrained the distances between the heavy atoms 
involved in the hydrogen bonds corresponding to the G-C pairs, using harmonic potentials.
Additional details of the simulations are given in the SI.

For each system we ran $32$ independent simulations, each approximately $1.0$-$1.5$ \textmu s long.
After an initial energy minimization using a steepest descent algorithm, $32$ independent simulations were initialized with random seeds
and simulated for $100$ ps at $T=400$ K, then equilibrated for additional $100$ ps at $T=300$ K.
The final configurations were used as starting points for the production runs.
The simulations using the DESRES and AMBER force fields were performed starting from exactly the same conformations.
Frames were stored for later analysis every $100$ ps.
The minimum distance observed between solute atoms from periodic images was 1.55 nm (DESRES simulations).
Stacking interactions were analyzed by using both the stacking score \cite{condon2015stacking} and the so-called G-vectors introduced in Ref.~\onlinecite{bottaro2014role}.
We analyzed the trajectories using Barnaba~\cite{bottaro2018analyze} and MDTraj.\cite{McGibbon2015MDTraj}

\subsection{Core Markov state model combined with density-based clustering}\label{sec:msm}
MSMs have been successfully applied to the study of many biomolecular
systems (see, \emph{e.g.}, Refs.~\onlinecite{pinamonti2016predicting,huang2009rapid,plattner2017complete,shukla2014activation,kohlhoff2014cloud,sadiq2012kinetic}).
The idea underlying an MSM is to reduce the complexity of a simulation by partitioning
the phase space into discrete microstates via a clustering algorithm.
The transition probabilities between these microstates can be then computed counting the transitions observed in the MD trajectories.
A possible approach to compute these probabilities is
the so-called ``transition-based-assignment'' or ``coring'',
first proposed in ref.~\onlinecite{buchete2008coarse}
and further analyzed in ref.~\onlinecite{schutte2011markov}.
The idea is to define a collections of ``core sets'', 
i.e. metastable regions of the phase space, 
which are not required to be in contact among each other.
A transition between states $A$ and $B$ is counted only when a trajectory goes
from the core region of $A$ ($\mathcal{C}_{A}$) to the core region of $B$ ($\mathcal{C}_B$)
without passing through any other core region.
Then the system will be considered in state $B$ until it goes back to $\mathcal{C}_A$
or reaches a third core region, independently of how many times it exits and
re-enters in $\mathcal{C}_B$ before reaching a new state.

  The fundamental step of this approach is to start with a good definition of metastable core sets.
This requirement is usually in contrast with the
fact that, when studying the dynamics of a complex biomolecule, no prior knowledge of the 
free-energy landscape of the system is available.
Therefore, in order to successfully apply this method it is necessary to extract this information 
from the simulation data, preprocessing the trajectories in order to identify different states 
and define realistic core regions.
A smart way to do this is to make use of a density-based clustering algorithm to 
separate the MD data set into a collection of clusters and identify
the core regions of these clusters as the regions with higher density.\cite{lemke2016density}

To construct core-based MSMs we proceed as follows.
We started by describing the system
using the same set of coordinates that we employed in a previous work:\cite{pinamonti2016predicting}
i) G-vectors (4D vectors connecting the nucleobases ring centers, as described in Ref.~\onlinecite{bottaro2014role}),
ii) the sine and cosine of backbone dihedrals, sugar ring torsional angles, and glycosidic torsional angles.
The dimensionality of this input was then reduced using
time-lagged independent component analysis
(TICA) \cite{perez2013identification}
with a lag time of $5$ ns, and
data was projected on the slowest TICs 
using a kinetic map projection.\cite{noe2015kinetic}
Subsequently, we used the pointwise-adaptive k-free energy estimator (PAk) algorithm~\cite{rodriguez2018computing}
combined with the TWO-NN algorithm~\cite{facco2016two} to estimate the pointwise density in TICA space,
which was then used to cluster the data using
density peak clustering.\cite{rodriguez2014clustering,d2018automatic}
We defined the core of each cluster as the set of all points $i$ for which
$\rho_i/\rho_\textrm{MAX}>e^{-1}$, where $\rho_{\textrm{MAX}}$ is the maximum density
in the cluster. According with \citet{rodriguez2018computing},  this corresponds approximately 
to a maximum of $1$ $k_\textrm{B}T$ free-energy difference between the configurations included in the core set
and those belonging to the transition areas.
Finally, the MD trajectories were discretized by assigning each frame to the last core set visited, and the
resulting discrete trajectories were used to estimate a reversible MSM.\cite{bowman2009progress,trendelkamp2015estimation}
More details on the procedure are given in the SI.

This procedure leads to robust and reliable MSMs.
In order to validate this procedure we compared the results with those of a standard MSM
in which the phase space was discretized
using k-means clustering on TICA projected space, and a transition was counted
every time a trajectory jumped from one microstate to the other.
Results of this validation are given in Section~\ref{results:validation}.

Afterwards, a lag time $\tau=100$ ps was then used to construct a core-based MSM that approximates the
dynamics of the discretized system.
The quality of the Markovian approximation was tested by looking 
at the convergence of the implied timescales predicted by the MSM
 for increasing values of $\tau$ %
as described in Ref.~\onlinecite{swope2004describing}. %

The MSM construction and analysis was performed
using the software PyEMMA 2.2.\cite{scherer2015pyemma}
Density Peak clustering was performed using the code available at \url{https://github.com/alexdepremia/Advanced-Density-Peaks}.

\subsection{Classification of states}

In order to obtain an easy-to-interpret representation of the fraying kinetics,
we classified the microstates obtained with the MSM procedure in
different groups. The classification was performed using a number of structural
determinants, including root-mean-square deviation (RMSD) from native conformation \cite{kabsch1976solution} and stacking
score.\cite{condon2015stacking}

For each system,
microstates were grouped
into the following states:
\begin{itemize}
\item closed ($C$): canonical double helix, with both terminal bases in their native conformations,
  stacking on the adjacent G or C base, and forming pairing interactions between each other;
\item open ($O$): frayed structures, with broken pairing between the two terminal bases,
  which are both unstacked and freely moving;
\item 5'-open ($5P$): the base at the 5' terminal base is not forming any stacking and pairing interactions, while the base in 3' is still in its native conformation;
\item 3'-open ($3P$): same as $5P$, but inverting 5' and 3';
\item misfolded ($M$): the base on the 3' terminus is rotated by $180$ degrees and stacking ``upside down'' on its adiacent base;
\item undefined ($U$): all conformations not falling into the previous categories, including, among others, microstates where
  the base at the 5' terminus is rotated upside down,
  or configurations in which one of the two terminal bases stacks on the top of a base in the opposite strand.
\end{itemize}
Technical details of this classification are reported in the SI. 

\section{Results}
\subsection{Validation of the core-based MSM}\label{results:validation}

As a first step we performed a validation of the introduced MSM procedure
using core-sets obtained with the PAk algorithm and DP clustering.
In particular, we here analyzed trajectories reported
in a previous paper~\cite{pinamonti2016predicting} for RNA adenine di- and tri-nucleotides.
Details of this analysis are provided in Section~\ref{sec:msm} and in the SI.
\begin{figure*}
\centering
\includegraphics[width=0.85\textwidth]
{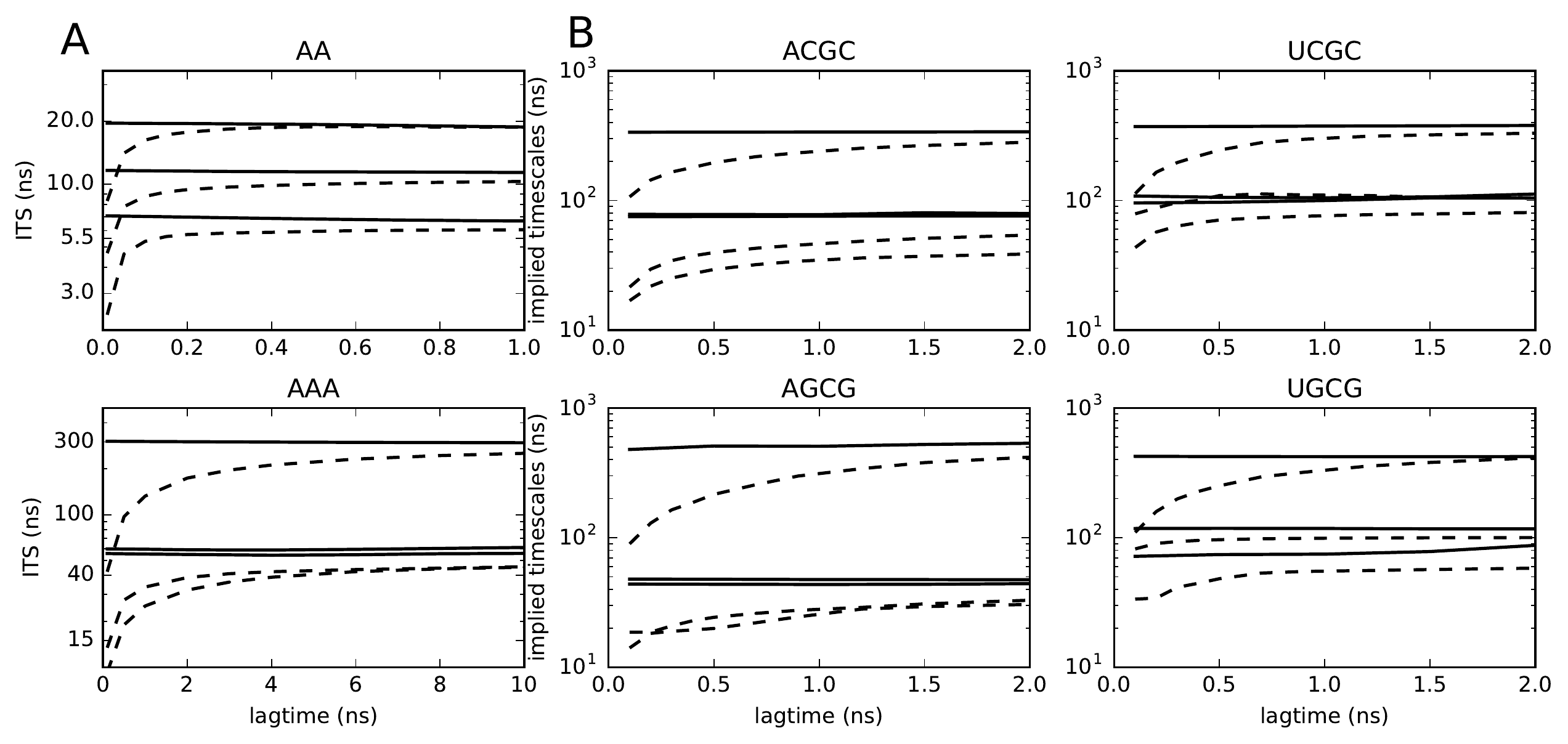}
\caption{Implied timescales of the MSM for different systems, as a function of the lag time.
  The three slowest timescales obtained with the core-based MSM (continuous lines) are compared with a standard MSM with k-mean clustering (dashed lines).
  Panel A shows the validation on the adenine di- and trinucleotide simulations taken from Ref.~\onlinecite{pinamonti2016predicting}.
  Panel B reports the results for the 4 RNA duplexes studied in this work.}
\label{fig:its}
\end{figure*}
Figure~\ref{fig:its}A reports a comparison between the results obtained with a core-set MSM
and those obtained with a standard MSM approach, as described in the Methods section.
Specifically, the timescales as a function of the
lag time $\tau$ are shown, and we can see that the core-based MSM lead to timescales
fully compatible with the standard approach.
Strikingly, the timescales are basically independent of the chosen
lag time, showing that this procedure is extremely robust and allows the selection of a
relatively short lag time for the MSM construction.
We also notice (See Tab.~SI~1) that the number of clusters resulting from the 
DP clustering is consistently smaller than the number of microstates that are required for a
good discretization using k-means.

\subsection{Energetic and kinetic analysis}

\begin{table}
      \caption{Thermodynamic and kinetic properties obtained from the MSMs
      of the 4 RNA duplexes, specifically:
      Free-energy difference between $O$ and $C$ states ($\Delta F_{C\rightarrow O}$)
      negative experimental stabilization of each duplex
      by the terminal base pair
      computed with nearest neighbors parameters ($-\Delta F_{\textrm{stab}}$) ;\cite{xia1998thermodynamic}
      MFPTs from $C$ to $O$, and viceversa;
      slowest implied timescale ($t_1$) obtained from the MSM.
      Energies are expressed in kcal/mol.
      Notice that the negative experimental stabilization is by construction expected to be smaller than the
      stacking energy (see text for discussion).
      Uncertainties were estimated by using a Jackknife procedure.
}

\centering
    \begin{tabular}{ccccccc}

\multirow{2}{*}{Seq.} &\multicolumn{1}{c}{$\Delta F_{C\rightarrow O}$ } & \multicolumn{1}{c}{$-\Delta F_{\textrm{stab}}$ } & &\multicolumn{2}{c}{MFPT (\textmu s)} & \multirow{2}{*}{$t_1$ (ns)}\\
      \cline{2-3}\cline{5-6}
 & MSM & Experiment & &$C \to O$  &$O \to C$ &\\
\hline
ACGC & $3.7\pm 0.05$ & $2.0$ && $10 \pm1$&$0.28\pm0.01$ &$336\pm12$\\ 
AGCG & $3.1\pm 0.15$ & $1.8$ && $10 \pm4$&$0.47\pm0.03$ &$479\pm31$\\ 
UCGC & $4.6\pm 0.02$ & $2.1$ && $52 \pm6$&$0.36\pm0.05$ &$371\pm57$\\ 
UGCG & $2.9\pm 0.07$ & $1.8$ && $10 \pm1$&$0.29\pm0.01$ &$424\pm21$\\  
\hline %
\end{tabular}
\label{tab:fray}
\end{table}
After being validated, the method is used to analyze large scale simulations of 4 short duplexes,
consisting in $32$ simulations, with a total simulation time of $35$ to $54$ \textmu s  for each sequence (see Tab.~SI~1 for details).
Two different force fields were employed. We here report results using the DESRES force field,\cite{tan2018rna}
whereas results using the standard AMBER force field are presented in
Section~\ref{results:comparison}.

From the equilibrium population of the microstates obtained from the MSM,
we computed the free-energy difference between the $O$ and $C$ states.
Tab.~\ref{tab:fray} reports the computed difference in free energy between the closed ($C$) state and the open ($O$) one.
The native structure is the most stable one, as expected.
The stability of the closed structure can be compared with thermodynamic experiments \cite{xia1998thermodynamic}
where the stabilization of a duplex due to the presence of an additional base pair is measured.
The negative of this number provides a lower bound for the stacking energy.
Indeed, by assuming that unstacked nucleobases do not form any interaction,
the contribution of an additional base pair to the stability of a duplex ($\Delta F_{\textrm{stab}}$)
can be approximated as
\begin{equation}
\Delta F_{\textrm{stab}} \approx \Delta F_{S\rightarrow U}^{ss_1} + \Delta F_{S\rightarrow U}^{ss_2}- \Delta F_{C\rightarrow O}~.
\end{equation}
Here $\Delta F_{S\rightarrow U}^{ss_1}$ and $\Delta F_{S\rightarrow U}^{ss_2}$ are the 
free-energy changes related to breaking a terminal stacking interaction in the two individual single strands.
Since single strands are expected to display a significant amount of stacking
\cite{bottaro2018conformational} these terms are expected to be positive,
so that $-\Delta F_{\textrm{stab}} < \Delta F_{C\rightarrow O}$.
The ranking of the four investigated systems is thus qualitatively consistent.

We also computed the stability of the intermediate states where only one of the two nucleobases
is open and the other is stacked (3'-open, $3P$, and 5'-open, $5P$, if the stacked nucleobase is at the 5' or 3' end of the helix, respectively).
The relative $\Delta F$, with respect to state $C$, of the $3P$ and $5P$ intermediate states are reported in Tab.~SI~3-6.
As expected, adenine (purine) terminal bases form stacking interactions that are stronger when compared with uracils (pyrimidines).
Moreover we find that the 5'-open states are on average
more stable than the 3'-open states, although the difference is modulated by the sequence.
In particular, the two stabilities are roughly comparable when the purine (A) is located at the 5'-end,
whereas the 5'-open state is significantly more stable when the purine is located at the 3'-end.
These results are qualitatively consistent with previous findings.\cite{colizzi2012rna}

Finally, states $U$ and $M$, where one or both nucleotides are not in their native structure nor unstacked,
appear with non-negligible population.
Their stabilities are significantly smaller than that of the native closed structure.
We are not aware of solution experiments that rule out these structures as possible alternatives.
Focusing on state $M$, which consists of a clearly defined ensemble of conformations,
we tried to search structures similar to these ones \cite{bottaro2014role} within the whole structural database using Barnaba.\cite{bottaro2018analyze}
Although fragments extracted from the database are expected to be highly biased due to their structural context and to
the variety of experimental conditions under which they were obtained, they were shown to agree with solution experiments to a significant extent both in
proteins \cite{best2006relation} and nucleic acids.\cite{bottaro2016rna}
A significant number of fragments with virtually identical base pairing
can be found (see Tab.~SI~6), suggesting that these misfolded structures are
plausible metastable states.

We then used the obtained MSMs to characterize the kinetics of fraying.
\begin{figure*}
\centering
\includegraphics[width=0.99\textwidth]
{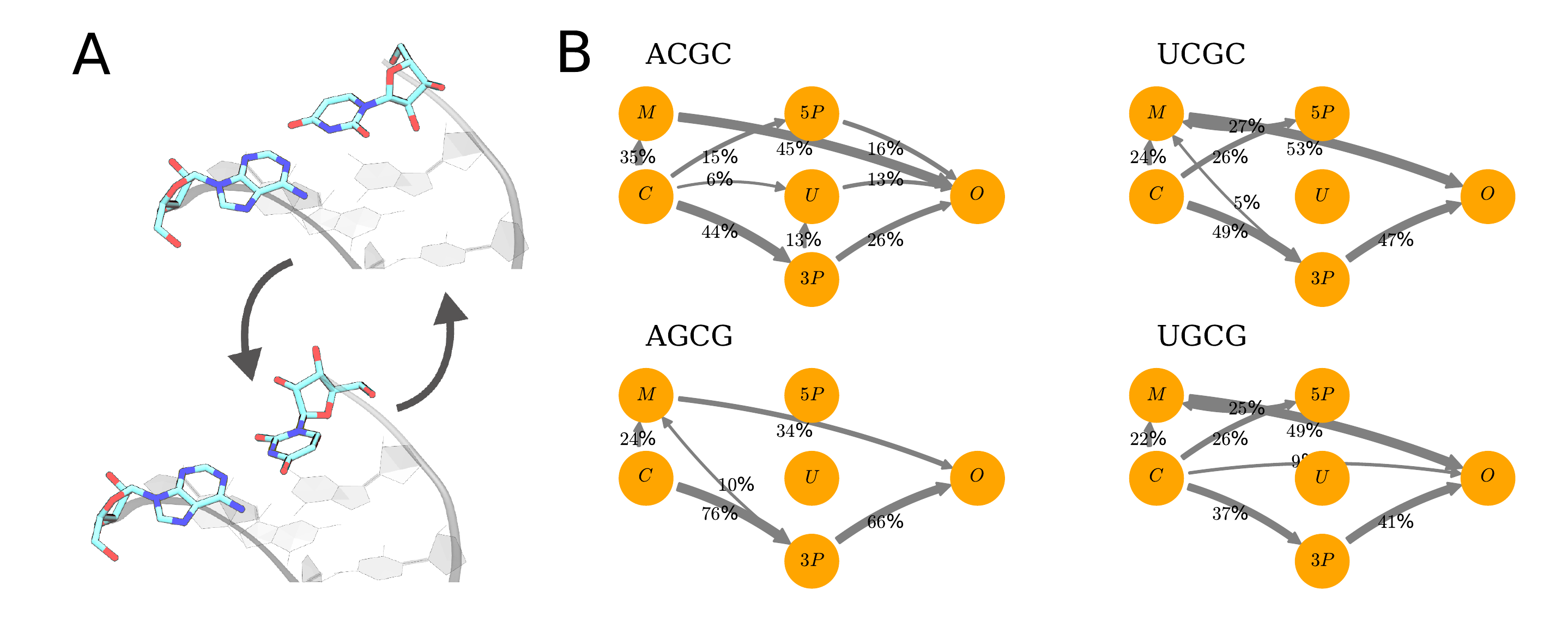}
\caption{Results of the MSMs of the four sequences, based on DESRES simulations.
  Panel A: Slowest process in the MSMs of the 4 duplexes.
Two structures from the simulations of ACGC are shown, as representatives for all sequences.
  Panel B: Flux of fraying trajectories computed from the four MSM by means of TPT.}
\label{fig:fray}
\end{figure*}
Interestingly, the slowest process always corresponds to the unstacking and rotation
of the nucleobase at the 3'-end (see Fig.~\ref{fig:fray}A), and thus represents the interconversion to the misfolded structure mentioned above.
The timescales of this process for the four different systems
are also reported in Table~\ref{tab:fray}.
We then computed the fraying kinetics for the four systems using transition-path-theory (TPT),\cite{weinan2006towards} in the formulation of MSMs. \cite{noe2009constructing}
The mean-first-passage time (MFPT) associated to the fraying transition for the four systems is reported in
Table~\ref{tab:fray}. This number is inversely proportional to the fraying rate. Interestingly, the MFPT for ACGC, AGCG, and UGCG are very similar ($10$ \textmu s) to each other,
while UGCG exhibits a much smaller fraying rate.
The MFPT for the inverse process, i.e. terminal pairing, is also reported in Table~\ref{tab:fray}.
We can see that this quantity shows a small dependence on the sequence
and is correlated with the timescale of the slowest process, i.e. the rotation of the 3' nucleobase,
showing that the state $M$ acts as a kinetic trap during terminal pairing.

We further investigated the mechanism of fraying, focusing on the first opening base.
Using TPT, we obtained the flux of fraying trajectories going from $C$ to $O$ through either $3P$ or $5P$.
Results are reported in Fig.~\ref{fig:fray}B.
The most likely path for fraying is, for the four investigated systems, the one through the $3P$ intermediate.
In other words, based on these results one would expect the nucleobase at the 3'-end to most likely break its stacking interation with the adjacent base
before the one at the 5'-end.
Interestingly, the most probable intermediate in the transition between $C$ and $O$ is always $3P$, even in sequences
UCGC and UGCG where
it is the \emph{least} stable one according to the free-energy analysis reported above.
This can be rationalized by the individual rates reported in Tab.~SI~3-6.
Estimated rates for transitions from $5P$ to $O$ are either null or very small.
This is a consequence of the fact that no or few transitions are observed in the MD simulation along this pathway.

\subsection{Comparison with AMBER force-field}\label{results:comparison}

We also analyzed an identical set of simulations performed using the latest
AMBER force field.\cite{banas2010performance}
Results are reported in Fig.~SI~2 and Tab.~SI~7.
These simulations resulted in a larger number
of non-canonical structures
when compared with those obtained in the simulations performed with the DESRES force field.
In particular, we observed a non-negligible population of so-called ladder-like structures
\citep{banas2010performance} (See Fig.~SI~2B).
Similar structures were also observed in previous works \cite{bergonzo2015highly,cunha2017unravelling} and
might be a consequence of both the short length of the duplex simulated here
and the presence of restraints on the base-pair distances in the duplex.

In addition, our simulations show a large population of misfolded structures.
See SI for more details about these structures.
In particular, for all the constructs except UCGC
the stability obtained of the misfolded structures was larger than that of the native structure.
Whereas these conformations are plausible metastable states
(see also Tab.~SI~6), their high populations make the results much
more difficult to interpret.

In general the closed state is less stable with respect to the open state. This can be seen both from the $\Delta F$, which are smaller in general, and from the shorter MFPTs (See Fig.~SI~2A)
Regarding the predicted fraying mechanics, most of the fraying pathways are going through the misfolded ($M$), the undefined ($U$), or the ladder-like ($L$) states. This makes it difficult to reach a definite conclusion regarding the mechanism. However, one can observe a general tendency for a mechanism where the 5' base opens before the 3' one, in contrast with what predicted from the simulations with DESRES force field.

\section{Discussion and Conclusions}

In this work, we developed a robust recipe for the construction of core-based MSMs and applied it to the characterization of fraying kinetics in RNA.
When compared with standard MSMs 
the core-based method enables to obtain MSMs with a limited number of microstates,
making the following analysis both clearer and more practical.
At the same time, the implied timescales are robustly estimated, even for very short values of
the lag time.
  One of the advantages of a short lag time is a statistically robust estimation of the MFPT, which is
  in general a challenging task for an MSM with a large lag time, due to the effect of recrossing
  events.
  
  We applied the introduced core-based MSM to study the thermodynamics and kinetics of base fraying,
  and analyzed the pathway followed during the process.
  We first focused on the free-energy difference between the native helical conformation
  and the frayed state using the DESRES force field.\cite{cornell1995second,perez2007refinement,tan2018rna}
This difference can be compared with the stabilization of a duplex resulting from the addition of an individual base pair,
as obtained from optical melting experiments.\cite{xia1998thermodynamic}
The ranking of the four analyzed sequences is qualitatively consistent with the experiments.
This result is by itself not obvious, given that
we are comparing systems with the same numbers of GC and AU pairs. In other words, the force field is capable to qualitatively capture
the difference between placing a purine or a pyrimidine on each of the two strands, and the interplay between the hydrogen bonds
formed in the first and in the second base pair of a helix.
The comparison is however only qualitative, since the experimental free energies report the difference in the stability
of two duplexes with a different number of pairs. By closing the thermodynamic cycle,
the experimental free energies should correspond to the difference between the stacking energy in a duplex and
the stacking energy in two separated single strands, that are not included here.
Previous works performing the full thermodynamic cycle using an older version of the AMBER force field
\cite{spasic2012amber} and the version used here \cite{sakuraba2015predicting} report agreement with
experimental free energies.
In addition, optical melting experiments are not sensitive to the precise structure
and only reports the overall  stability of a bimolecular complex that might originate from the combination of different structures.
It must be also observed that, although the DESRES force field is 
the only force field to date that was shown to be able to predict the folded structure
of RNA tetraloops including their signature interactions, \cite{tan2018rna}
its capability to reproduce experimentally observed non-canonical interactions 
has been recently questioned.\cite{kuhrova2018improving}

  We then focused on the kinetics of the fraying process, estimating the fraying rate and the
  weight of different pathways.
  The fraying rates of three of the four sequences are all around $10^5$ s$^{-1}$.
  The fourth sequence, UCGC, displays a 5-times slower fraying. This effect can be attributed
  to the larger stability of the UC and AG stacking interactions, also observed in the thermodynamic parameters.
The reported rates are in qualitative agreement with those measured using imino-proton exchanges.
\cite{snoussi2001imino}
We observe that rates for terminal pairing show a slight dependence with respect to the sequence.
The ratio between the lowest and highest rate is $\approx 1.7$, corresponding to a contribution
of $\approx 0.3$ kcal/mol to the sequence-dependence of the stability of individual base pairs.
In comparison, the ratio between the highest and the lowest unpairing rate is $\approx 5$,
corresponding to a contribution of $\approx 1$ kcal/mol.
This agrees with the common notion that the stability differences depend more on the off rates than on the on rates.
\cite{bloomfield2000nucleic}

  Interestingly, the fraying path with the largest flux always corresponds to a 3'-open intermediate, which is typically
  the least stable among the two intermediates.
  The opening of the 5' end likely results in a reclosure of the pair because the 3' dangling end stays stacked.
  The opening of the 3' end can result in complete opening because the 5’ dangle is less stable.
  Therefore, a 3' end opening, although less frequent, is more likely to proceed to complete opening.
  The 5'-open path has been proposed as the most likely one based on the frequency of 3'-dangling bases
  in crystal structures \cite{mohan2009mechanism} and on the relative stability of the two intermediates as computed by
  molecular simulations.\cite{colizzi2012rna}
  It is however important to underline that the results reported here might be affected by the choice of the force field.
  In particular, we are not aware of any validation of the kinetics reported so far for the DESRES RNA force field.

Lastly, we report a comparison with simulations performed with the standard AMBER force field.\cite{cornell1995second,perez2007refinement,banas2010performance}
Results are significantly different in the stability of the canonical duplexes, in the estimated fraying rates, and in the predicted pathways.
In particular, the AMBER force field generates a larger population of non-canonical, misfolded structures.
Whereas the observed ladder-like structures \cite{banas2010performance} might be an artifact related to the short length of the simulated helices,
the non-canonical interactions at the terminal bases have been already reported elsewhere in the context of longer duplexes \cite{zgarbova2014base}
and are probably intrinsically stabilized by this force field.
The relative population of these structures might change using recent corrections that aimed at providing a better balance between
important hydrogen bond interactions.\cite{kuhrova2018improving}
Overall, the presence of these structures makes the interpretation of the fraying process more difficult.

Interestingly, whereas the relative stability of the different intermediates
exhibits a similar trend between different force-fields,
the pathways of fraying are substantially different.
Indeed, it is known that the impact of force fields on kinetics can be large even when the
thermodynamic properties are similar.\cite{vitalini2015dynamic}
In this particular case, the different kinetics might be a consequence of different energetic barriers (e.g., in the backbone torsional angles)
that are difficult to validate experimentally. The possibility to have different force fields resulting in the same
native structure but predicting different pathways have been discussed for instance in the case of protein folding simulations.\cite{piana2011robust}

      In conclusion, we constructed a core-based MSM with
      the goal of reproducing the kinetic properties
      of the terminal base pair of an RNA double helix.
The introduced method
 makes it possible to obtain a robust estimation of rates 
      and MFPT between folded and frayed structures,
      to identify metastable states and their 
      relative stabilities,
      as well as the unzipping pathways followed by the system. 
Although the obtained rates are in qualitative agreement with experimental data,
the appearance of non-canonical structures and/or their excessive suppression make the preferential unzipping pathway
dependent on the chosen force field.

\begin{acknowledgements}
  G.P. and G.B. received support by the European Research Council, Starting Grant 306662.
  F.N. and F.P. acknowledge funding from European Research Commission (ERC StG 307494 "pcCell"
  and ERC CoG 772230 "ScaleCell") and Deutsche Forschungsgemeinschaft (SFB1114/C03 and SFB1114/A03).
  F.P. acknowledges funding from the Yen Post-Doctoral Fellowship in Interdisciplinary Research and from the National Cancer Institute of the National Institutes of Health (NIH) through Grant CAO93577.
  We thank Alessandro Laio, Maria d'Errico, and Elena Facco, for their help in the clustering analysis.
  We are grateful to Sandro Bottaro for many helpful advices and for interesting scientific discussions.
  We also thank the anonymous reviewers whose comments have greatly improved this manuscript.
\end{acknowledgements}

\end{document}